\documentclass[prd,tightenlines]{revtex4}
\usepackage{epsfig}
\usepackage{graphics}
\usepackage{latexsym}
\usepackage{amsmath}
\usepackage{amssymb}
\usepackage{rotating}

\newcommand{\gev}{\,{\rm GeV}}

\begin{document}

\title{Nucleon magnetic form factors with non-local chiral effective Lagrangian}

\author{P. Wang}

\affiliation{Institute of High Energy Physics, CAS, P. O. Box
918(4), Beijing 100049, China}

\affiliation{Theoretical Physics Center for Science Facilities,
CAS, Beijing 100049, China}

\begin{abstract}

Chiral perturbation theory is a powerful method to investigate the hadron properties. We apply the non-local
chiral effective Lagrangian to study nucleon magnetic form factors. The octet and decuplet intermediate states
are included in the one loop calculation. With the modified propagators and non-local interactions,
the loop integral is convergent. The obtained proton and neutron magnetic form factors are both reasonable
up to relatively large $Q^2$.

\end{abstract}


\maketitle

\section{Introduction}

The study of electromagnetic properties of hadrons has attracted a lot of interest for many years.
Though QCD is the fundamental theory to describe the strong interaction, it is difficult
to apply it directly to study the hadron properties due to the non-perturbative property.
There are many phenomenological models based on hadron or quark level such as
cloudy bag model \cite{Lu:1997sd}, the constituent quark model
\cite{Berger:2004yi,JuliaDiaz:2003gq}, the $1/N_c$ expansion
approach \cite{Buchmann:2002et}, the perturbative chiral quark model
\cite{Cheedket:2002ik}, non-local quark meson coupling model \cite{Faessler}, the extended vector meson dominance model
\cite{Williams:1996id}, the quark-diquark model
\cite{Hellstern:1995ri} and the Schwinger-Dyson formalism
\cite{Oettel:1998bk,Alkofer:2004yf,Eichmann:2007nn}, etc.

As well as the above model calculations, there are also many lattice-QCD studies of the
electromagnetic form factors. Significant efforts to probe baryon
electromagnetic structure in lattice QCD have been driven by the
Adelaide group \cite{Zanotti:2003gc,Boinepalli:2006xd}, the Cyprus
group \cite{Alexandrou:2006ru}, and the QCDSF
\cite{Gockeler:2003ay,Gockeler:2006ui,Gockeler:2007hj} and LHP
Collaborations \cite{Edwards:2005kw,Alexandrou:2005fh}.
Though lattice QCD is the most rigorous approach, most
quantities are simulated with large quark ($\pi$) mass because of
the computing limitations.
Therefore, it is necessary to extrapolate the lattice data to the
physical $\pi$ mass.

Chiral perturbation theory ($\chi$PT) is a systematic tool in studying hadron physics and
the Lagrangian is well defined and based on the chiral symmetry which is the same as QCD.
Various formulations of $\chi$PT have also been widely applied to study the hadron properties.
\cite{Puglia:2000jy,Fuchs:2003ir,Kubis:2000aa,Kubis:2000zd}.
To deal with the divergence of the loop integral, most formulations of $\chi$PT are based on dimensional
or infrared regularization. With dimensional regularization, it has
been observed that expansions in $\chi$PT are consistent with
experimental results up to $Q^2 \simeq 0.1\gev^2$
\cite{Fuchs:2003ir}. Extensions of $\chi$PT to explicitly incorporate
vector mesons have been demonstrated to improve the applicability to
$Q^2 \simeq 0.4\gev^2$ \cite{Schindler:2005ke}.

An alternative regularization method, namely finite-range-regularization (FRR) is
widely applied in the chiral extrapolation of the lattice data.
A ultraviolet regulator reflects the structure of hadrons is introduced in the loop integral.
FRR effective field theory (EFT) was first applied in the extrapolation of the nucleon mass and
magnetic moments~\cite{Leinweber:1998ej,Leinweber0,Leinweber1}. The
remarkably improved convergence properties of the FRR expansion mean
that lattice data at large pion masses can be described very well and
the nucleon mass obtained at the physical pion mass compared favorably
with the experimental value.  Later, the FRR method was applied to
extrapolate the vector meson mass, magnetic moments, magnetic form
factors, strange form factors, charge radii, first moments of GPDs,
etc.\ \cite{Alton,Armour,Young,Wang1,Leinweber:2004tc,Leinweber:2006ug,Wang2,Wang3,Wang4,Wang5}.

Therefore, to study the hadron properties at relatively large $Q^2$ and pion mass. It is important
to consider the size effect of the hadrons. In the previous work, we proposed a new quantization
condition, i.e. solid quantization \cite{solid1,solid2}. With the solid quantization, the modified propagators of the hadrons as well as the
non-local interaction are obtained. Compared with the FRR EFT, the ultraviolet regulator can be derived
from the Lagrangian. In this paper, we will apply the non-local Lagrangian in the heavy baryon formalism
to study the nucleon magnetic form factors up to relatively large $Q^2$.
The paper is organized as follows. In section II, we briefly introduce the chiral Lagrangian which will be used
in our loop calculation. The nucleon magnetic form factor is derived in section III. Numerical results are presented
in section IV. Finally, section V is a summary.

\section{Chiral Lagrangian}

There are many papers which deal with heavy baryon chiral
perturbation theory -- for details see, for example, Refs.
\cite{Jenkins:1990jv,Bernard:1992qa,Bernard:2007zu}. For
completeness, we briefly introduce the formalism in this section. In
the heavy baryon chiral perturbation theory, the lowest chiral
Lagrangian for the baryon-meson interaction which will be used in
the calculation of the nucleon magnetic moments, including the octet
and decuplet baryons, is expressed as

\begin{eqnarray}
{\cal L}_v &=&i{\rm Tr}\bar{B}_v(v\cdot {\cal D}) B_v+2D{\rm
Tr}\bar{B}_v S_v^\mu\{A_\mu,B_v\} +2F{\rm Tr}\bar{B}_v
S_v^\mu[A_\mu,B_v]
\nonumber \\
&& -i\bar{T}_v^\mu(v\cdot {\cal D})T_{v\mu} +{\cal C}(\bar{T}_v^\mu
A_\mu B_v+\bar{B}_v A_\mu T_v^\mu),
\end{eqnarray}
where $S_\mu$ is the covariant spin-operator defined as
\begin{equation}
S_v^\mu=\frac i2\gamma^5\sigma^{\mu\nu}v_\nu.
\end{equation}
Here, $v^\nu$ is the nucleon four velocity (in the rest frame, we
have $v^\nu=(1,0)$). D, F and $\cal C$ are the coupling constants.
The chiral covariant derivative $D_\mu$ is written as $D_\mu
B_v=\partial_\mu B_v+[V_\mu,B_v]$. The pseudoscalar meson octet
couples to the baryon field through the vector and axial vector
combinations
\begin{equation}
V_\mu=\frac12(\zeta\partial_\mu\zeta^\dag+\zeta^\dag\partial_\mu\zeta),~~~~~~
A_\mu=\frac12(\zeta\partial_\mu\zeta^\dag-\zeta^\dag\partial_\mu\zeta),
\end{equation}
where
\begin{equation}
\zeta=e^{i\phi/f}, ~~~~~~ f=93~{\rm MeV}.
\end{equation}
The matrix of pseudoscalar fields $\phi$ is expressed as
\begin{eqnarray}
\phi=\frac1{\sqrt{2}}\left(
\begin{array}{lcr}
\frac1{\sqrt{2}}\pi^0+\frac1{\sqrt{6}}\eta & \pi^+ & K^+ \\
\pi^- & -\frac1{\sqrt{2}}\pi^0+\frac1{\sqrt{6}}\eta & K^0 \\
K^- & \bar{K}^0 & -\frac2{\sqrt{6}}\eta
\end{array}
\right).
\end{eqnarray}
$B_v$ and $T^\mu_v$ are the velocity dependent new fields which are
related to the original baryon octet and decuplet fields $B$ and
$T^\mu$ by
\begin{equation}
B_v(x)=e^{im_N \not v v_\mu x^\mu} B(x),
\end{equation}
\begin{equation}
T^\mu_v(x)=e^{im_N \not v v_\mu x^\mu} T^\mu(x).
\end{equation}
In the chiral $SU(3)$ limit, the octet baryons will have the same
mass $m_B$. In our calculation, we use the physical masses for
baryon octets and decuplets. The explicit form of the baryon octet
is written as
\begin{eqnarray}
B=\left(
\begin{array}{lcr}
\frac1{\sqrt{2}}\Sigma^0+\frac1{\sqrt{6}}\Lambda &
\Sigma^+ & p \\
\Sigma^- & -\frac1{\sqrt{2}}\Sigma^0+\frac1{\sqrt{6}}\Lambda & n \\
\Xi^- & \Xi^0 & -\frac2{\sqrt{6}}\Lambda
\end{array}
\right).
\end{eqnarray}
For the baryon decuplets, there are three indices, defined as
\begin{eqnarray}
T_{111}=\Delta^{++}, ~~ T_{112}=\frac1{\sqrt{3}}\Delta^+, ~~
T_{122}=\frac1{\sqrt{3}}\Delta^0, \\ \nonumber T_{222}=\Delta^-, ~~
T_{113}=\frac1{\sqrt{3}}\Sigma^{\ast,+}, ~~
T_{123}=\frac1{\sqrt{6}}\Sigma^{\ast,0}, \\ \nonumber
T_{223}=\frac1{\sqrt{3}}\Sigma^{\ast,-}, ~~
T_{133}=\frac1{\sqrt{3}}\Xi^{\ast,0}, ~~
T_{233}=\frac1{\sqrt{3}}\Xi^{\ast,-}, ~~ T_{333}=\Omega^{-}.
\end{eqnarray}

The octet, decuplet and octet-decuplet transition magnetic moment
operators are needed in the one loop calculation of nucleon magnetic
form factors. The baryon octet magnetic Lagrangian is written as:
\begin{equation}\label{lomag}
{\cal L}=\frac{e}{4m_N}\left(\mu_D{\rm Tr}\bar{B}_v \sigma^{\mu\nu}
\left\{F^+_{\mu\nu},B_v\right\}+\mu_F{\rm Tr}\bar{B}_v
\sigma^{\mu\nu} \left[F^+_{\mu\nu},B_v \right]\right),
\end{equation}
where
\begin{equation}
F^+_{\mu\nu}=\frac12\left(\zeta^\dag F_{\mu\nu}Q\zeta+\zeta
F_{\mu\nu}Q\zeta^\dag\right).
\end{equation}
$Q$ is the charge matrix $Q=$diag$\{2/3,-1/3,-1/3\}$. At the lowest
order, the Lagrangian will generate the following nucleon magnetic
moments:
\begin{equation}\label{treemag}
\mu_p=\frac13\mu_D+\mu_F,~~~~~~ \mu_n=-\frac23\mu_D.
\end{equation}

The decuplet magnetic moment operator is expressed as
\begin{equation}
{\cal L}=-i\frac{e}{m_N}\mu_C
q_{ijk}\bar{T}^\mu_{v,ikl}T^\nu_{v,jkl} F_{\mu\nu},
\end{equation}
where $q_{ijk}$ and $q_{ijk}\mu_C$ are the charge and magnetic
moment of the decuplet baryon $T_{ijk}$. The transition magnetic
operator is
\begin{equation}
{\cal L}=i\frac{e}{2m_N}\mu_T F_{\mu\nu}\left(\epsilon_{ijk}Q^i_l
\bar{B}^j_{vm} S^\mu_v T^{\nu,klm}_v+\epsilon^{ijk}Q^l_i
\bar{T}^\mu_{v,klm} S^\nu_v B^m_{vj}\right).
\end{equation}
In Ref.~\cite{Durand1}, the authors used $\mu_u$, $\mu_d$ and
$\mu_s$ instead of the $\mu_C$ and $\mu_T$. For the particular
choice, $\mu_s=\mu_d=-\frac12 \mu_u$, one finds the following
relationship:
\begin{equation}
\mu_D=\frac32 \mu_u, ~~~ \mu_F=\frac23 \mu_D, ~~~ \mu_C=\mu_D, ~~~
\mu_T=-4\mu_D.
\end{equation}
In our numerical calculations, the above formulas are used and
therefore all baryon magnetic moments are related to one parameter,
$\mu_D$.

The above local Lagrangian need to be replaced by the non-local form if we take the size of the hadrons into account.
The gauge invariant Lagrangian of the first term of Eq.(1) can be written as \cite{solid2}
\begin{equation}
\int d^3 ai\bar{\psi}'(t, \vec{x}+\frac{\vec{a}}{2})v^\mu\cdot ({\cal D}_\mu - ie_{eff} A_\mu(x)) \psi'(t,\vec{x}-\frac{\vec{a}}{2}) F(\vec{a}),
\end{equation}
where
\begin{equation}
\bar{\psi}'(t, \vec{x}+\frac{\vec{a}}{2}) = \bar{\psi}(t, \vec{x}+\frac {\vec{a}}{2}) e^{ie_{eff}^j I(\vec{x}+\vec{a}/2,\vec{x})},
\end{equation}
\begin{equation}
\psi'(t, \vec{x}-\frac{\vec{a}}{2}) =  e^{-ie_{eff}^j I(\vec{x}-\vec{a}/2,\vec{x})} \psi(t,\vec{x}-\frac {\vec{a}}{2}),
\end{equation}
where $e_{eff}^j$ is expressed as $\frac{e_j\Psi(\vec{a})}{F(\vec{a})}$ and $e_j$ is the charge of hadron $j$.
Similarly, the local nucleon-meson interaction can be changed to be a non-local interaction for
non-point particles. For example,
\begin{equation}
{\rm Tr}\bar{B}_v S_v^\mu \partial_\mu \phi B_v \Rightarrow \int d^3a \int d^3b{\rm Tr}\left[\bar{B}'_v(x+\frac {\vec{a}}{2}) S_v^\mu \partial_\mu \phi'(x+\vec{b})B'_v(x-\frac {\vec{a}}{2})\right]\Phi_B(\vec{a})\Phi_M(\vec{b}).
\end{equation}
where
\begin{equation}
\phi'(\vec{x}+\vec{b}) = e^{-ie_{eff}^j I(\vec{x}+\vec{b},\vec{x})}\phi(\vec{x}+\vec{b}).
\end{equation}
with $I(\vec{y},\vec{x}) = \int_{\vec{x}}^{\vec{y}} dz_\mu A^{\mu}(z)$. In the above equations, $\tilde{\Phi}_B(\vec{p}^2)$ and $\tilde{\Phi}_M(\vec{p}^2)$ are the Fourier transformation of
$\Phi_B(\vec{a})$ and $\Phi_M(\vec{b})$ which are related to the free nucleon and meson wave function.

As in Ref.~\cite{solid2}, due to the non-point property of hadrons, in the
heavy baryon formalism, the propagators of the octet and
decuplet baryon, $j$ can be written as
\begin{equation}
\frac {i\tilde{\Phi}_B(\vec{p}^2)} {v\cdot k-\delta^{jN}+i\varepsilon} ~~{\rm and}~~ \frac
{iP^{\mu\nu}\tilde{\Phi}_B(\vec{p}^2)} {v\cdot k-\delta^{jN}+i\varepsilon},
\end{equation}
where $P^{\mu\nu}$ is $v^\mu v^\nu-g^{\mu\nu}-(4/3)S_v^\mu S_v^\nu$.
$\delta^{ab}=m_b-m_a$ is the mass difference of between the two
baryons. The propagator of meson $j$ ($j=\pi$, $K$, $\eta$) is expressed
as
\begin{equation}
\frac {i\tilde{\Phi}_M(\vec{k}^2)} {k^2-M_j^2+i\varepsilon}.
\end{equation}

\section{Magnetic Form Factors}

\begin{figure}[tbp]
\begin{center}
\includegraphics[scale=0.45]{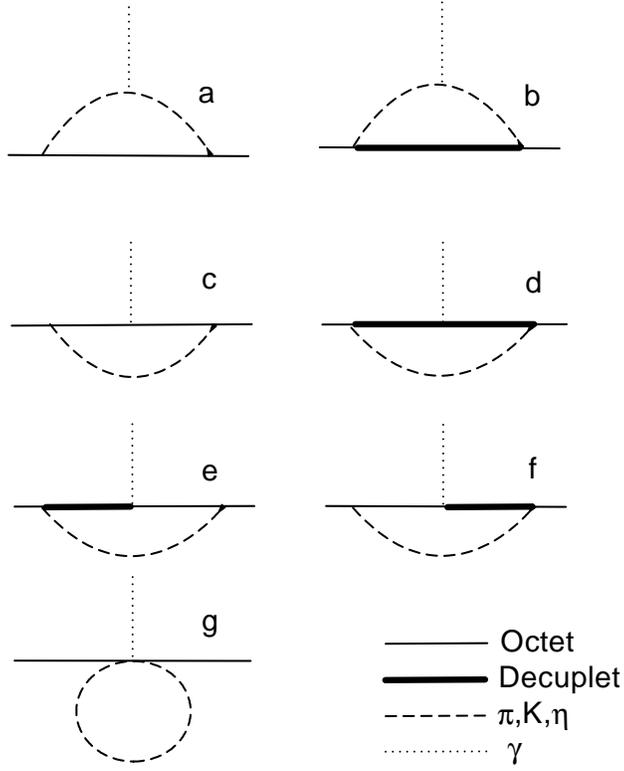}
\caption{Feynman diagrams for the nucleon magnetic form factor.}
\label{diagrams}
\end{center}
\end{figure}

In the heavy baryon formalism, the nucleon form factors are
defined as:
\begin{equation}
<B(p^\prime)|J_\mu|B(p)>=\bar{u}(p^\prime)\left\{v_\mu
G_E(Q^2)+\frac{i\epsilon_{\mu\nu\alpha\beta}v^\alpha S_v^\beta
q^\nu}{m_N}G_M(Q^2)\right\}u(p),
\end{equation}
where $q=p^\prime-p$ and $Q^2=-q^2$. According to the Lagrangian,
the one loop Feynman diagrams which contribute to the nucleon
magnetic moments are plotted in Fig.~1. Fig.~1a and Fig.~1b
provide the leading order contribution while the other diagrams
give the next to leading order contribution. The contributions to nucleon
magnetic form factors of Fig.~1a are expressed as
\begin{equation}\label{p1a}
G_M^{p(1a)}=\frac{m_N(D+F)^2}{8\pi^3\Lambda^2}I_{1\pi}^{NN}
+\frac{m_N(D+3F)^2I_{1K}^{N\Lambda}+3m_N(D-F)^2I_{1K}^{N\Sigma}}{{48\pi^3\Lambda^2}},
\end{equation}
\begin{equation}\label{n1a}
G_M^{n(1a)}=-\frac{m_N(D+F)^2}{8\pi^3\Lambda^2}I_{1\pi}^{NN}
+\frac{m_N(D-F)^2}{8\pi^3\Lambda^2}I_{1K}^{N\Sigma}.
\end{equation}
The integration $I_{1j}^{\alpha\beta}$ is expressed as
\begin{equation}
I_{1j}^{\alpha\beta}=\int d\overrightarrow{k}\frac{k_y^2
W_1(\omega_j(\overrightarrow{k}+\overrightarrow{q}/2)
+\omega_j(\overrightarrow{k}-\overrightarrow{q}/2)+\delta^{\alpha\beta})}
{A_j^{\alpha\beta}},
\end{equation}
where
\begin{eqnarray}
A_j^{\alpha\beta}&=&\omega_j(\overrightarrow{k}+\overrightarrow{q}/2)
\omega_j(\overrightarrow{k}-\overrightarrow{q}/2)
(\omega_j(\overrightarrow{k}+\overrightarrow{q}/2)+\delta^{\alpha\beta})
\nonumber \\
&&(\omega_j(\overrightarrow{k}-\overrightarrow{q}/2)+\delta^{\alpha\beta})
(\omega_j(\overrightarrow{k}+\overrightarrow{q}/2)+\omega_j(\overrightarrow{k}-\overrightarrow{q}/2)).
\end{eqnarray}
$\omega_j(\overrightarrow{k})=\sqrt{m_j^2+\overrightarrow{k}^2}$ is
the energy of the meson $j$. $W_1$ is the additional function related to $\tilde{\Phi}_B$ and $\tilde{\Phi}_M$, expressed as
\begin{equation}
W_1 = \tilde{\Phi}((\frac{\vec{q}}{2})^2)\tilde{\Phi}^2(\vec{k}^2)\tilde{\Phi}^2((\vec{k}+\frac{\vec{q}}{2})^2)
\tilde{\Phi}^2((\vec{k}-\frac{\vec{q}}{2})^2)\tilde{\Phi}((\frac{\vec{k}}{2}+\frac{\vec{q}}{4})^2)\tilde{\Phi}((\frac{\vec{k}}{2}-\frac{\vec{q}}{4})^2).
\end{equation}
The first terms in Eqs.\ (\ref{p1a}) and (\ref{n1a}) come from the $\pi$
meson cloud contribution. The second terms come from the K meson cloud
contribution. This diagram was studied in our
previous paper \cite{Wang1,Wang5} where the regulator is introduced ``by hand".
Here, the function $W_1$ is obtained from the modified propagator and the non-local Lagrangian.
The modified baryon and meson propagators give the factor
$\tilde{\Phi}_B(\vec{k}^2)\tilde{\Phi}_M((\vec{k}+\frac{\vec{q}}{2})^2)
\tilde{\Phi}_M((\vec{k}-\frac{\vec{q}}{2})^2)$. The non-local baryon-meson interaction
provides the factor $\tilde{\Phi}_M((\vec{k}+\frac{\vec{q}}{2})^2)
\tilde{\Phi}_M((\vec{k}-\frac{\vec{q}}{2})^2)\tilde{\Phi}_B((\frac{\vec{k}}{2}+\frac{\vec{q}}{4})^2)\tilde{\Phi}_B((\frac{\vec{k}}{2}-\frac{\vec{q}}{4})^2)$,
while the non-local photon-meson interaction provides the factor $\tilde{\Phi}_M(\vec{k}^2)$.
The factor $\tilde{\Phi}_B((\frac{\vec{q}}{2})^2)$ is from the external free nucleon.
For simplification, we have chosen the correlation function $\tilde{\Phi}_B(\vec{p}^2)=\tilde{\Phi}_M(\vec{p}^2)=\tilde{\Phi}(\vec{p}^2)$.

Fig.~1b is the same as Fig.~1a but the intermediate
states are decuplet baryons. Their contributions to the magnetic form
factors are expressed as
\begin{equation}
G_M^{p(1b)}=\frac{m_N{\cal C}^2}{36\pi^3\Lambda^2}I_{1\pi}^{N\Delta}
-\frac{m_N{\cal C}^2}{144\pi^3\Lambda^2}I_{1K}^{N\Sigma^\ast},
\end{equation}
\begin{equation}
G_M^{n(1b)}=-\frac{m_N{\cal C}^2}{36\pi^3\Lambda^2}I_{1\pi}^{N\Delta}
-\frac{m_N{\cal C}^2}{72\pi^3\Lambda^2}I_{1K}^{N\Sigma^\ast}.
\end{equation}
The contributions to the form factors from Fig.~1c are expressed as
\begin{eqnarray} \nonumber
G_M^{p(1c)}&=&\frac{(D+F)^2(\mu_D-\mu_F)}{192\pi^3\Lambda^2}I_{2\pi}^{NN}
-\frac{1}{192\pi^3\Lambda^2}\left[(D-F)^2(2\mu_F+\mu_D)
I_{2K}^{N\Sigma}-(\frac D3+F)^2\mu_DI_{2K}^{N\Lambda}
\right. \\
&& \left.
-(D-F)(\frac{2D}3+2F)\mu_DI_{5K}^{N\Lambda\Sigma}\right]
-\frac{(\frac
D3-F)^2(\mu_D+3\mu_F)}{192\pi^3\Lambda^2}I_{2\eta}^{NN},
\end{eqnarray}
\begin{eqnarray}\nonumber
G_M^{n(1c)}&=&-\frac{(D+F)^2\mu_F}{96\pi^3\Lambda^2}I_{2\pi}^{NN}
-\frac{1}{192\pi^3\Lambda^2}\left[(D-F)^2(\mu_D-2\mu_F)I_{2K}^{N\Sigma}-(\frac D3+F)^2
\mu_DI_{2K}^{N\Lambda}
\right. \\
&& \left.
+(\frac{2D}3+2F)(D-F)\mu_DI_{5K}^{N\Lambda\Sigma}\right]
+\frac{(\frac D3-F)^2\mu_D}{96\pi^3\Lambda^2} I_{2\eta}^{NN},
\end{eqnarray}
where
\begin{equation}
I_{2j}^{\alpha\beta}=\int d\overrightarrow{k}\frac{k^2 W_2}
{\omega_j(\overrightarrow{k})(\omega_j(\overrightarrow{k})+\delta^{\alpha\beta})^2},
\end{equation}
\begin{equation}
I_{5j}^{\alpha\beta\gamma}=\int d\overrightarrow{k}\frac{k^2 W_2}
{\omega_j(\overrightarrow{k})(\omega_j(\overrightarrow{k})+\delta^{\alpha\beta})
(\omega_j(\overrightarrow{k})+\delta^{\alpha\gamma}))}.
\end{equation}
The function of $W_2$ is expressed as
\begin{equation}
W_2 = \tilde{\Phi}((\frac{\vec{q}}{2})^2)\tilde{\Phi}^4(\vec{k}^2)\tilde{\Phi}((\vec{k}+\frac{\vec{q}}{2})^2)
\tilde{\Phi}((\vec{k}-\frac{\vec{q}}{2})^2)\tilde{\Phi}((\frac{\vec{k}}{2}+\frac{\vec{q}}{2})^2)\tilde{\Phi}((\frac{\vec{k}}{2}-\frac{\vec{q}}{2})^2).
\end{equation}
The magnetic moments of nucleon in the chiral limit, expressed in
terms of $\mu_D$ and $\mu_F$, are used in the one loop calculations.

The contributions to the form factors of Fig.~1d are expressed as
\begin{equation}
G_M^{p(1d)}=\frac{5{\cal C}^2\mu_C}{162\pi^3\Lambda^2}I_{2\pi}^{N\Delta}
+\frac{5{\cal C}^2\mu_C}{1296\pi^3\Lambda^2}I_{2K}^{N\Sigma^\ast},
\end{equation}
\begin{equation}
G_M^{n(1d)}=-\frac{5{\cal C}^2\mu_C}{648\pi^3\Lambda^2}I_{2\pi}^{N\Delta}
-\frac{5{\cal C}^2\mu_C}{1296\pi^3\Lambda^2}I_{2K}^{N\Sigma^\ast}.
\end{equation}
Fig.~1e and Fig.~1f give the following contributions to the form
factors:
\begin{equation}
G_M^{p(1e+1f)}=\frac{(D+F){\cal C}\mu_T}{108\pi^3\Lambda^2}I_{3\pi}^{N\Delta}
+\frac{5(D-F){\cal
C}\mu_T}{864\pi^3\Lambda^2}I_{5K}^{N\Sigma\Sigma^\ast}
+\frac{(D+3F){\cal
C}\mu_T}{864\pi^3\Lambda^2}I_{5K}^{N\Lambda\Sigma^\ast},
\end{equation}
\begin{equation}
G_M^{n(1e+1f)}=-\frac{(D+F){\cal C}\mu_T}{108\pi^3\Lambda^2}I_{3\pi}^{N\Delta}
+\frac{(D-F){\cal C}\mu_T}{864\pi^3\Lambda^2}I_{5K}^{N\Sigma\Sigma^\ast}
-\frac{(D+3F){\cal C}\mu_T}{864\pi^3\Lambda^2}I_{5K}^{N\Lambda\Sigma^\ast},
\end{equation}
where
\begin{equation}
I_{3j}^{\alpha\beta}=\int d\overrightarrow{k}\frac{k^2 W_2}
{\omega_j(\overrightarrow{k})^2(\omega_j(\overrightarrow{k})+\delta^{\alpha\beta})}.
\end{equation}

The total nucleon magnetic form factors can be written as
\begin{equation}
G_M^p(Q^2)=ZG_M^{p0} + \sum_{k=a}^f G_M^{p(1k)}(Q^2),
\end{equation}
\begin{equation}
G_M^n(Q^2)=ZG_M^{n0} + \sum_{k=a}^f G_M^{n(1k)}(Q^2),
\end{equation}
where $G_M^{p0}$ and $G_M^{n0}$ are the tree level magnetic form factors expressed
as
\begin{equation}
G_M^{p0}=(\frac13\mu_D+\mu_F)\tilde{\Phi}((\frac{\vec{q}}{2})^2),~~~~ G_M^{n0}=-\frac23\mu_D\tilde{\Phi}((\frac{\vec{q}}{2})^2).
\end{equation}
$Z$ is the wave function renormalization constant, expressed as
\begin{equation}
Z = 1 - \frac{3(D+F)^2}{64 \pi^3 \Lambda^2} \int d^3 k \frac{
\vec{k}^2 W_3}{\omega^3(\vec{k})}- \frac{{\cal C}^2}{24
\pi^3 \Lambda^2} \int d^3 k \frac{\vec{k}^2
W_3}{\omega(\vec{k})(\omega(\vec{k})+\delta)^2},
\end{equation}
where $W_3$ is expressed as
\begin{equation}
W_3 = \frac{\tilde{\Phi}((\frac{\vec{q}}{2})^2)}{\tilde{\Phi}(Q^2=0)}W_1(Q^2=0)
\end{equation}

Here we did not include the tadpole contribution. In fact, the tadpole contribution
is highly suppressed due to the finite size of the hadrons. For example, for the proton magnetic form
factors, the tadpole contribution is written as
\begin{equation}
G_M^{p(tad)} = -\frac{(\mu_D+\mu_F)}{32\pi^3\Lambda^2}I_{4\pi},
\end{equation}
where
\begin{equation}
I_{4\pi}=\int d^3 k\frac{W_4}
{\omega_\pi(\vec{k})}.
\end{equation}
$W_4$ is expressed as
\begin{equation}
W_4 = \int \frac{d^3l}{(2\pi)^3}\tilde{\Phi}((\frac{\vec{q}}{2})^2)\tilde{\Phi}(\vec{k}^2)\tilde{\Phi}(\vec{l}^2)\tilde{\Phi}((\vec{k}+\vec{l})^2)\tilde{\Phi}((\vec{k}-\vec{l})^2).
\end{equation}
The function $\tilde{\Phi}(\vec{k}^2)$ is from the meson propagator, while the additional $l$ integral of $\tilde{\Phi}(\vec{l}^2)\tilde{\Phi}((\vec{k}+\vec{l})^2)\tilde{\Phi}((\vec{k}-\vec{l})^2)$
is due to the non-local four-particle interaction.
The tadpole contribution of Fig. 1g was not included in our previous work of nucleon magnetic form factors \cite{Wang5,Wang2014}.
In this work, numerical result shows that the tadpole contribution to the nucleon form factors is very small.
It is less than $1\%$ of nucleon magnetic moments. Therefore, it can be neglected naturely in our calculation.

\section{Numerical Results}

\begin{figure}[tbp]
\begin{center}
\includegraphics[scale=0.65]{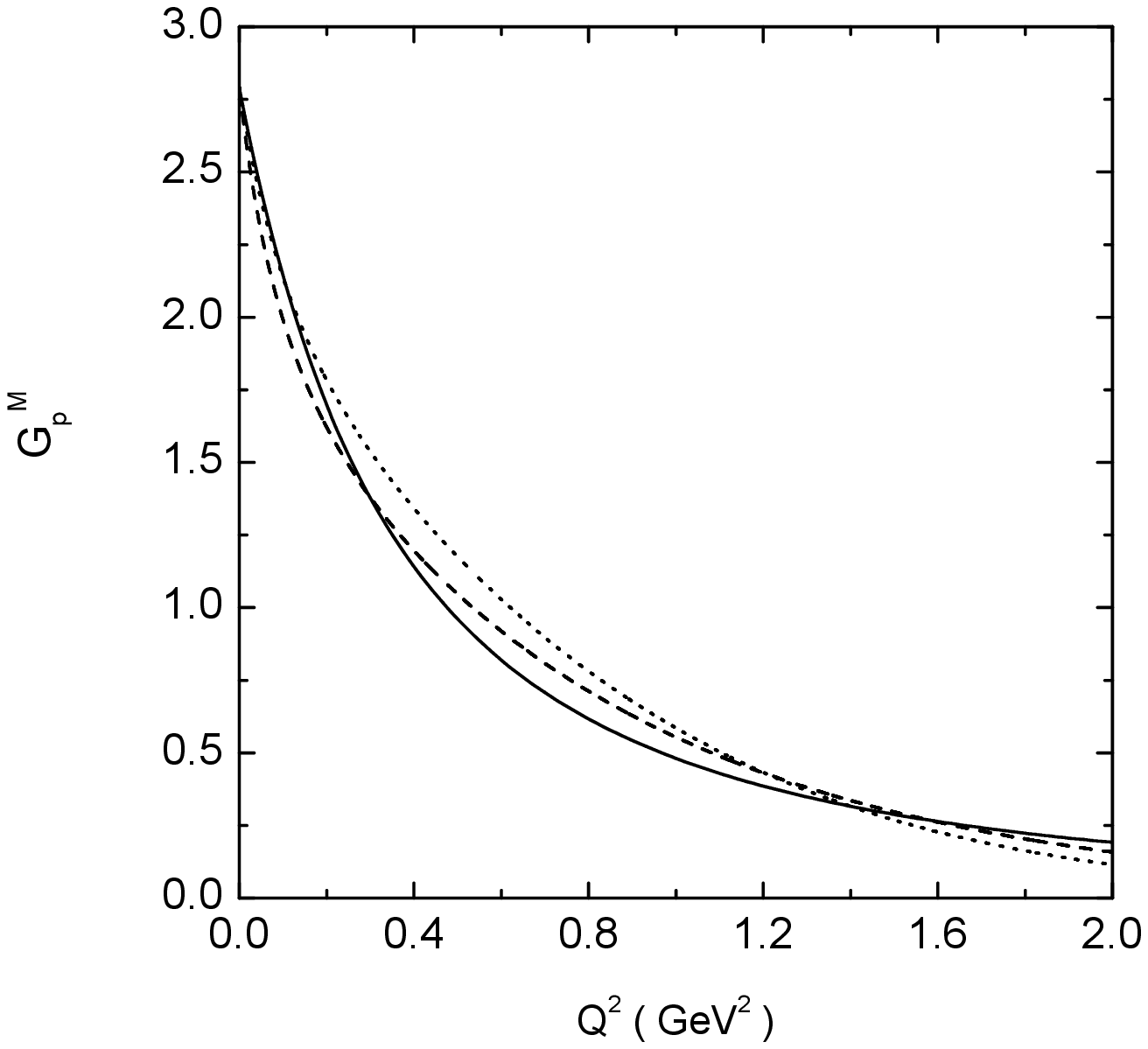}
\caption{Proton magnetic form factor versus momentum transfer $Q^2$. The solid line is for the empirical result. The dashed and dotted
lines are for the standard and modified Gauss functions, respectively.}
\label{diagrams}
\end{center}
\end{figure}

\begin{figure}[tbp]
\begin{center}
\includegraphics[scale=0.65]{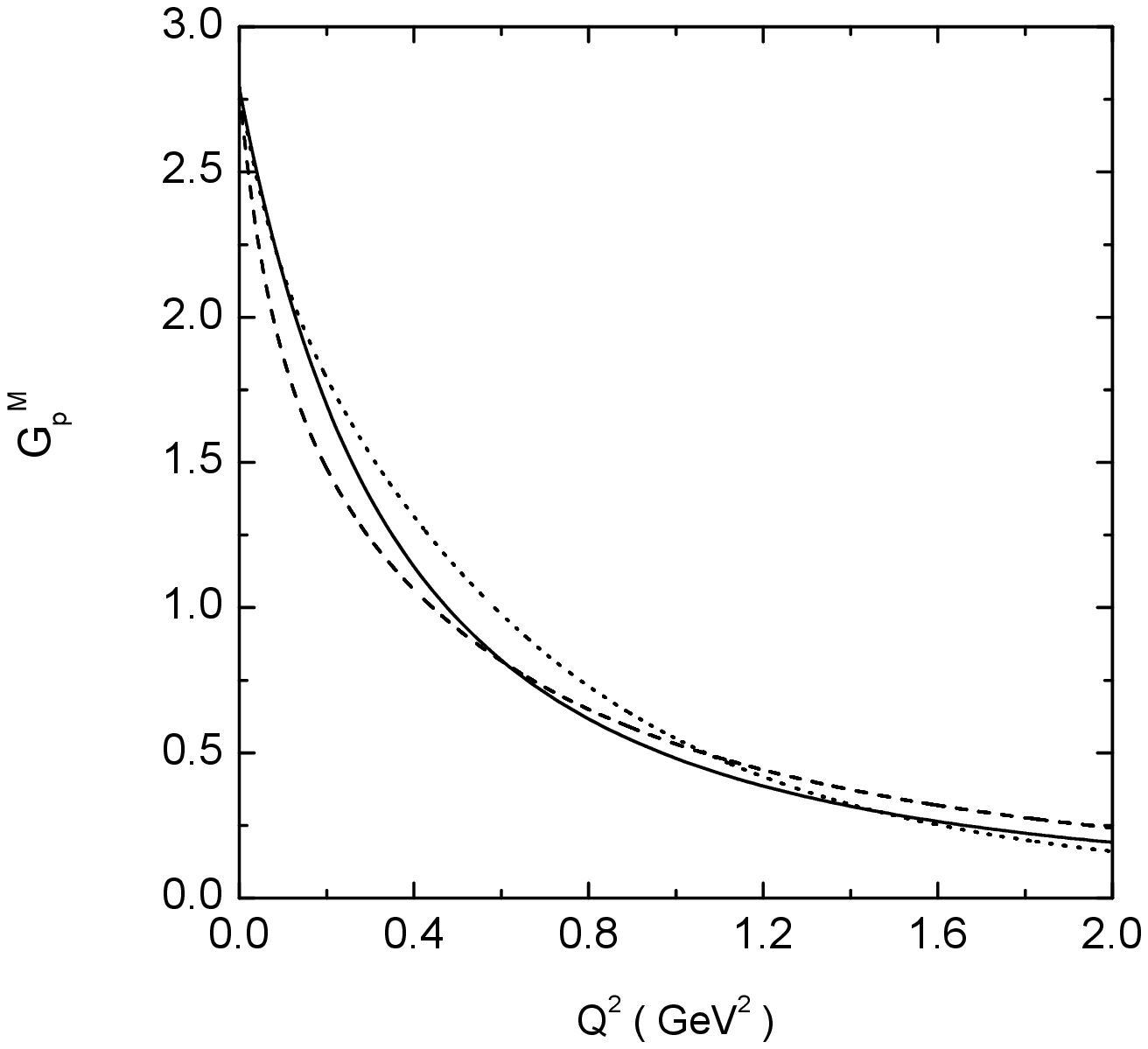}
\caption{Same as Fig.~ 2 except that the dashed and dotted lines are for the standard and modified dipole functions, respectively.}
\label{diagrams}
\end{center}
\end{figure}

In the numerical calculations, the parameters are chosen as $D=0.76$
and $F=0.50$ ($g_A=D+F=1.26$). The coupling constant ${\cal C}$ is
chosen to be $-1.2$ which is the same as Ref.\ \cite{Jenkins2}.
There is a parameter $\Lambda$ in the Lagrangian, whose value is $F_\pi$
for the local case. Here, the local interaction is modified to the
non-local one, $\Lambda$ is different from $F_\pi$ which will be
determined as a free parameter.

The low energy constant $\mu_D$ is chosen as our previous paper \cite{Wang1,Wang5},
i.e. $\mu_D$ is 2.40 and 2.05 for proton and neutron, respectively.
For the correlator, four kinds of functions are tested.
First we calculate the proton magnetic form factor with standard Gauss function
\begin{equation}
\tilde{\Phi}_1(\vec{k}^2) = exp\{-(\frac{|\vec{k}|}{\Lambda'})^2\}
\end{equation}
and modified Gause function
\begin{equation}
\tilde{\Phi}_2(\vec{k}^2) = exp\{-(\frac{|\vec{k}|}{\Lambda'})^{2.5}\}.
\end{equation}
In the above functions, $\Lambda{'}$ is a parameter. Therefore, there are
two free parameters $\Lambda$ and $\Lambda{'}$ need to be determined.
They are determined by the experimental proton and neutron magnetic moments.

\begin{figure}[t]
\begin{center}
\includegraphics[scale=0.65]{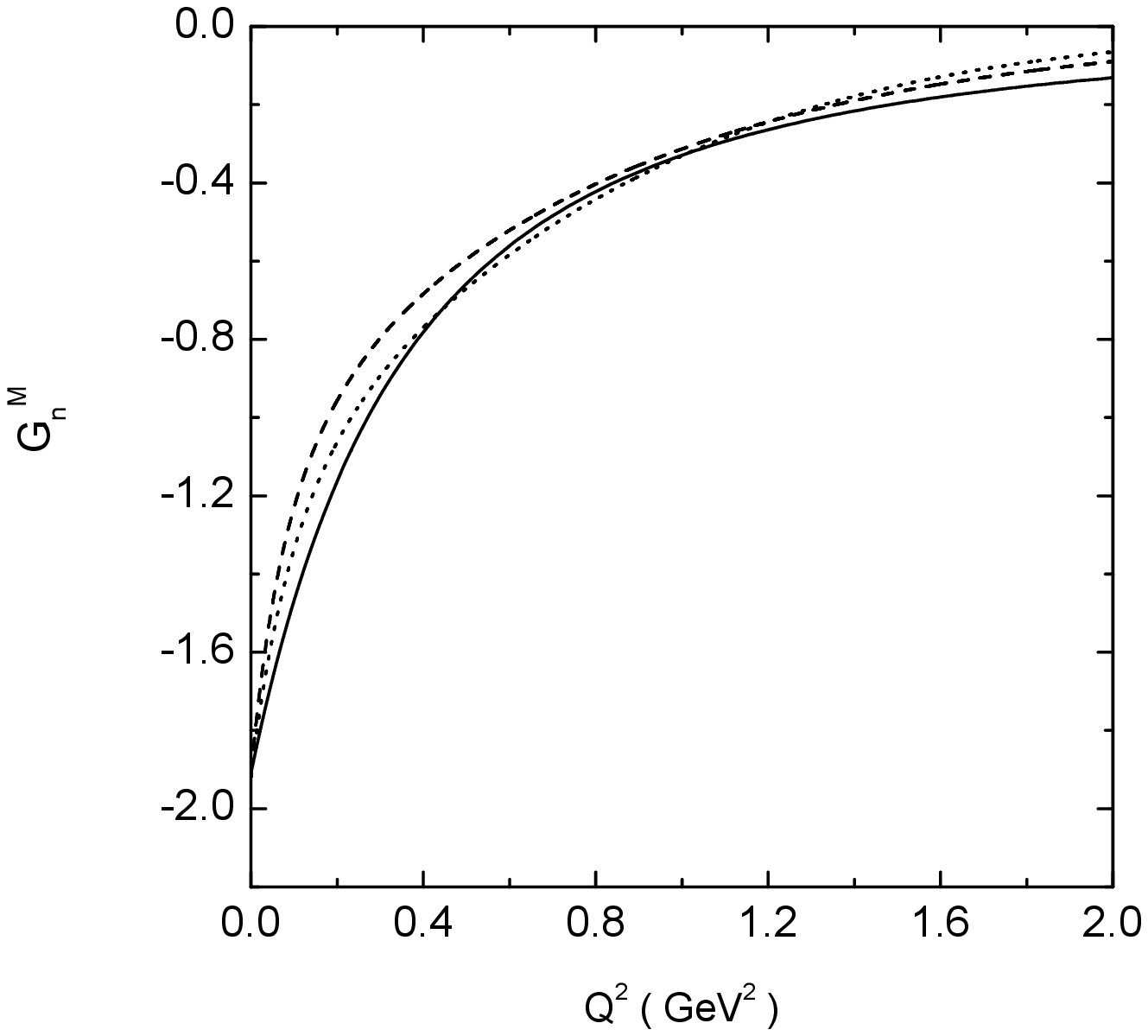}
\caption{Same as Fig.~2 but for neutron magnetic form factor.}
\label{diagrams}
\end{center}
\end{figure}

\begin{figure}[t]
\begin{center}
\includegraphics[scale=0.65]{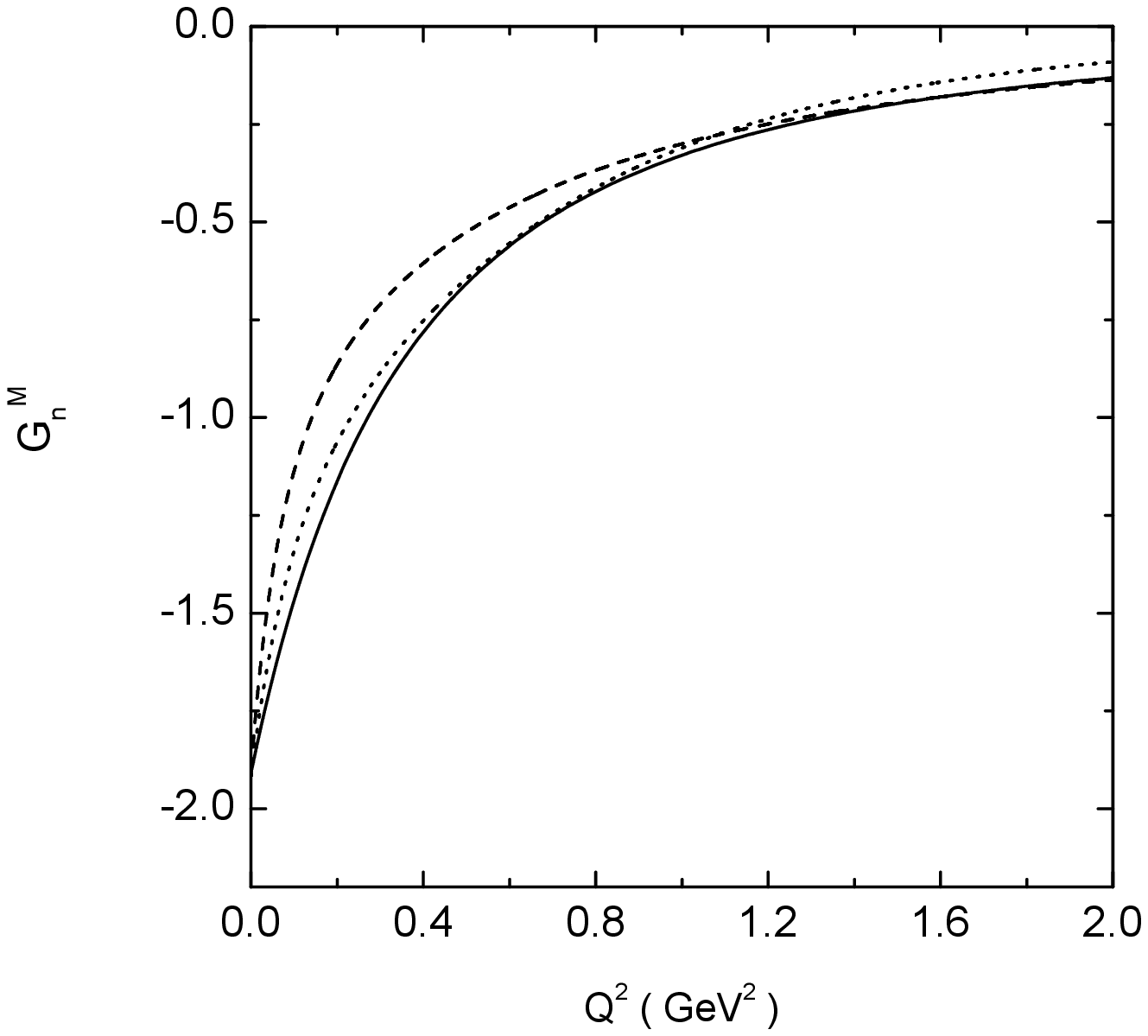}
\caption{Same as Fig.~3 but for neutron magnetic form factor.}
\label{diagrams}
\end{center}
\end{figure}

The proton magnetic form factor $G_M^p(Q^2)$ versus $Q^2$ is plotted in Fig.~2.
The dashed and dotted lines are for the standard and modified Gauss functions, respectively.
The solid line is for the empirical parametrization $G^p_M(Q^2) = 2.79/(1 +Q^2/0.71$ GeV$^2)^2$.
From the figure, one can see that the calculated proton magnetic form factor is comparable with
the experimental data for both correlators up to $Q^2= 2$ GeV$^2$.
For the standard Gauss function, the proton magnetic form factor at low $Q^2$ drops faster than the modified Gauss function.
resulting a larger magnetic radius. The radii are $1.12$ fm$^2$ and $0.72$ fm$^2$, respectively.
The radius from the modified Gauss function is close to the experimental value which can also be seen
from Fig.~2 since at low $Q^2$, the dotted line is much closer to the empirical line.

We also tried the standard dipole function
\begin{equation}
\tilde{\Phi}_3(\vec{k}^2) = 1/(1+(\frac{|\vec{k}|}{\Lambda'})^2)^2
\end{equation}
and modified dipole function
\begin{equation}
\tilde{\Phi}_4(\vec{k}^2) = 1/(1+(\frac{|\vec{k}|}{\Lambda'})^3)^2.
\end{equation}
The numerical results are plotted in Fig.~3. Same as Fig.~2 except the dashed and dotted lines are for the standard and modified dipole
functions, respectively. In these cases, the proton magnetic form factors can be described well up to $2$ GeV$^2$
as well. At low $Q^2$, the modified dipole function behaves better which gives a better magnetic radius.

The neutron magnetic form factors for the Gauss-type functions are shown in Fig.~4.
The solid, dashed and dotted lines are results of the empirical, the standard Gauss and modified Gauss functions, respectively.
For the calculation of neutron form factor, all the parameters are kept same as in the proton case.
From the figure, one can see the calculated neutron magnetic form factor is comparable with
the experimental data up to 2 GeV$^2$. At low energy transfer, the neutron magnetic form factor
increases faster than the empirical data for both cases which means the calculated magnetic radii are larger than
the experimental data. The magnetic radius of neutron obtained with the modified Gauss function is about 1 fm
which is comparable with the experimental data 0.87 fm.

The neutron magnetic form factors for the dipole-type functions are shown in Fig.~5.
Again, the calculated form factor is comparable with the empirical data up to $Q^2=2$ GeV$^2$.
The radius obtained from modified dipole function is also about 1 fm which better than that
for the standard dipole function.

\begin{figure}[tbp]
\begin{center}
\includegraphics[scale=0.65]{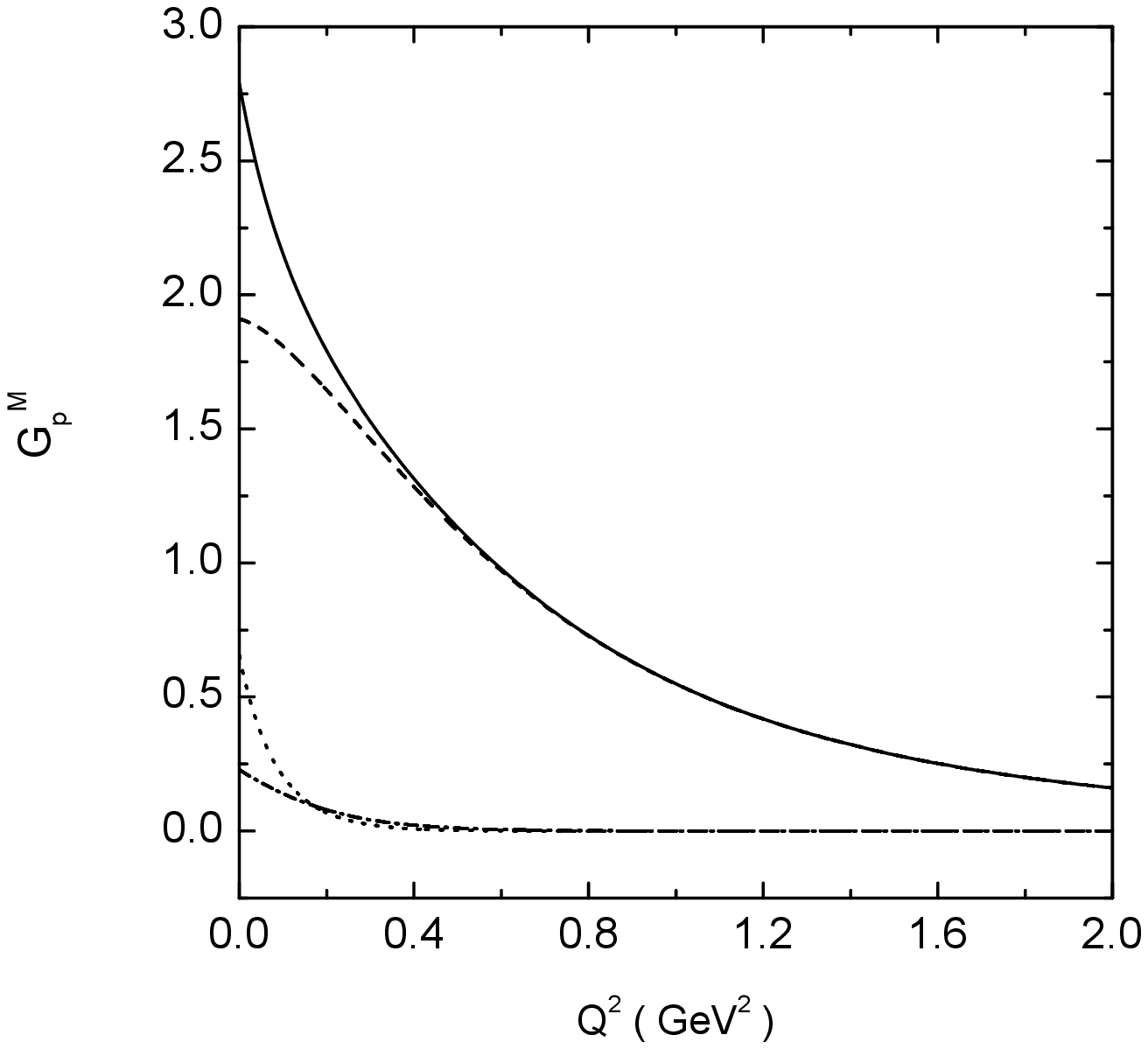}
\caption{Proton magnetic form factor versus momentum transfer $Q^2$. The solid, dashed, dotted and dash-dotted lines
are for the total, tree-level, leading order and next to leading order contribution, respectively}
\label{diagrams}
\end{center}
\end{figure}

\begin{figure}[tbp]
\begin{center}
\includegraphics[scale=0.65]{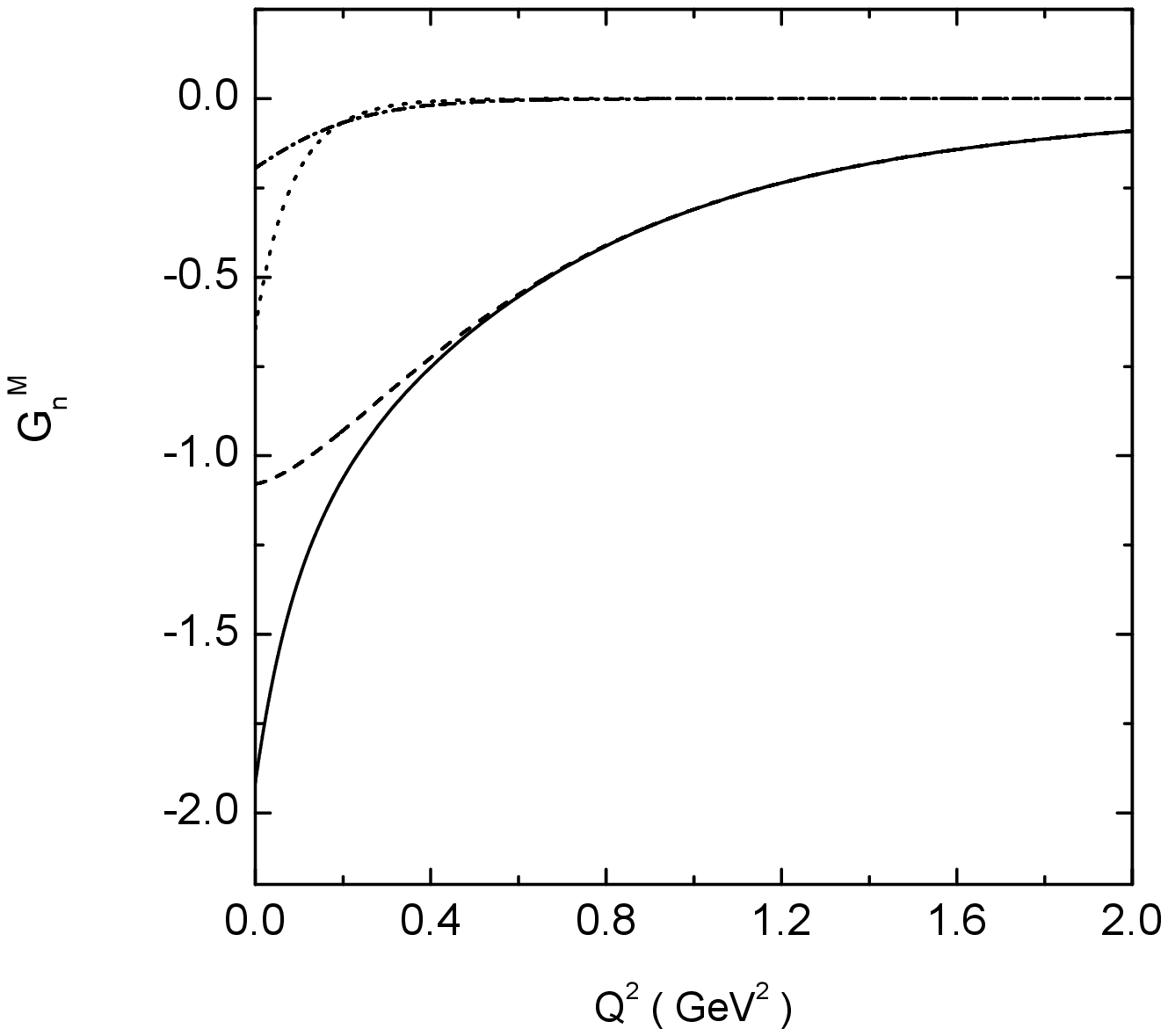}
\caption{Same as Fig.~6 but for neutron magnetic form factor.}
\label{diagrams}
\end{center}
\end{figure}

In Fig.~6, we plot the contributions to the proton magnetic form factor separately.
The solid, dashed, dotted and dash-dotted lines are for the total, tree-level, leading order and next to leading order
contribution, respectively. From the figure, one can see that loop and tree-level
contribute about $30\%$ and $70\%$ to the proton magnetic moment. With the increasing $Q^2$,
the tree-level (3-quark core) contribution is dominant. When $Q^2$ is larger than about 0.4
GeV$^2$, there is no visible contribution from the meson loop. This result can be easy
understood from the meson cloud picture where the 3-quark core of nucleon is surrounded
by the meson cloud. At low energy transfer, the meson cloud is important to the nucleon
form factors, especially to the nucleon radii. When $Q^2$ becomes large, the photon will
detect the 3-quark core. Therefore, the 3-quark core contribution is dominant to the proton
magnetic form factor at large $Q^2$.

Similarly, the total, tree-level, leading order and next to leading order
contributions to the neutron magnetic form factor are plotted in Fig.~7.
Again, one can see that the loop contribution is important for the neutron magnetic
moment and radius. At large $Q^2$, the 3-quark core contribution is dominant.
No visible contribution to the neutron magnetic form factor from meson loop
when $Q^2$ is larger than about 0.4 GeV $^2$.

\begin{table}
\caption{.}
\begin{ruledtabular}
\begin{tabular}{cccccc|ccccc}
& & Proton  & & & & & Neutron
\\ \hline
 & $\Lambda$ (GeV)  & $\Lambda'$ (GeV) & $\mu_D$ & $\mu_p$ & $r_M^2$ $(fm^2)$ & $\Lambda$ (GeV)  & $\Lambda'$ (GeV) & $\mu_D$ & $\mu_n$ & $r_M^2$ $(fm^2)$
\\ \hline
Set 1 & 0.032 & 0.45 &  2.40 & 2.79 & 1.12 & 0.032 & 0.45 & 2.03 & $-1.91$ & 1.47   \\
Set 2 & 0.041 & 0.47 &  2.40 & 2.79 & 0.72 & 0.041 & 0.47 & 2.03 & $-1.92$ & 1.02  \\
Set 3 & 0.026 & 0.52 &  2.40 & 2.79 & 1.40 & 0.026 & 0.52 & 2.03 & $-1.90$ & 1.82   \\
Set 4 & 0.042 & 0.52 &  2.40 & 2.79 & 0.73 & 0.042 & 0.52 & 2.03 & $-1.92$ & 1.04  \\
\end{tabular}
\end{ruledtabular}
\end{table}

\section{Summary}

We studied the nucleon magnetic form factors with the non-local chiral Lagrangian.
The one loop integral is not divergent due to
the correlation function. The baryon octets and decuplets are include in the intermediate states.
There are only two free parameters $\Lambda$ and $\Lambda'$ which is determined by
the experimental nucleon moments. The parameters and results for the four kinds of functions
are summarized in Table I. Set 1, set 2, set 3 and set 4 are for the standard Gauss,
modified Gauss, standard dipole and modified dipole function, respectively.

The contribution to the form factors from tadpole diagram is very small and can be neglected naturely.
We tested four kinds of correlation functions, i.e. Gauss-type and dipole-type functions.
The nucleon form factors can be described well up to $Q^2=2$ GeV$^2$ for all these four
kind of functions. The magnetic radii obtained from modified Gauss and dipole functions
are comparable with the empirical data. No visible contribution to the nucleon magnetic
form factors from the meson loop when $Q^2$ is larger than about 0.4 GeV$^2$.

\section*{Acknowledgments}

This work is supported in part by DFG and NSFC (CRC 110) and by the
National Natural Science Foundation of China (Grant No. 11035006).

\end{document}